# Design guidelines for efficient plasmonic solar cells exploiting the trade-off between scattering and metallic absorption

Xiaofeng Li[1,2,*], Nicholas P. Hylton[3], Vincenzo Giannini[3], Ned J. Ekins-Daukes[3] and Stefan A. Maier[3]

[1]*College of Physics, Optoelectronics and Energy & Collaborative Innovation Center of Suzhou Nano Science and Technology, Soochow University, Suzhou 215006, China.*
[2]*Key Lab of Advanced Optical Manufacturing Technologies of Jiangsu Province & Key Lab of Modern Optical Technologies of Education Ministry of China, Soochow University, Suzhou 215006, China.*
[3]*Blackett Laboratory, Department of Physics, Imperial College London, London SW7 2AZ, UK*
[*]*Corresponding author: xfli@suda.edu.cn*

**Abstract**

We report on the role of plasmonic resonances in determining the delicate balance between scattering and absorption of light in nanometric particle arrays applied to the front surface of solar cells. Strong parasitic absorption is shown to be dependent upon the excitation of localized surface plasmon resonances and prohibits efficient scattering into the underlying semiconductor. Via detailed analytical and numerical investigations we obtain the dependence of scattering and absorption in nanoparticles upon their complex refractive index. These results provide an insight into the optimum material properties required to minimize parasitic optical absorption, while maintaining high scattering cross-section efficiency, thus providing a general design guideline for efficient light trapping with scattering nanoparticles. The work is extended to include comprehensive optoelectronic simulations of plasmonic solar cells in which the scattering metals are made from either Au, Ag or Al. We show that Al particles provide the closest approximation to the optimized particle refractive index and therefore exhibit the smallest parasitic absorption and correspondingly lead to the greatest solar cell efficiency enhancements. Indeed, for the Al particles we report a full-band enhancement of external quantum efficiency over the reference device.

## 1. Introduction

Incorporation of plasmonic nanostructures into thin-film solar cells (SCs) has been extensively discussed in recent years [1–3], due to the capability of confining solar energy into a subwavelength region so that a long optical path in physically thin photoactive layers can be realized [4]. There have been a large variety of theoretical and experimental implementations to justify the performance enhancement of SCs arising from plasmonics [5–25], based on inorganic or organic photosensitive materials. The plasmonic nanostructures can also be configured into different locations in the cell, e.g., at the entrance side for resonantly enhanced forward scattering [5–19], inside the active layer to excite the localized surface plasmons [20–22], or the rear side for light reflection and near-field confinement [7, 23–25].

Among these systems, nanoscale light trapping by putting nanostructures on top of the cell is a classic design that attracted initial interest. This approach is still being paid considerable attention [5–19], since 1) the fabrication of nanostructures does not intervene in the growth of the photo-absorbing layers (i.e., PN junctions) and 2) the metallic layer can help to reduce the series resistance of the cell. The primary motivation of this design is to introduce metallic nanoparticles (NPs, normally gold or silver) on the front to generate (plasmonic) resonantly enhanced forward scattering, leading to a longer optical path length in the cell and hence increased absorption [2, 3]. The resonances of NPs are tunable by controlling the particle shape, size, mutual coupling, or the surrounding environment [17]. Performance enhancements have indeed been observed from both simulations and experiments [2, 3, 5–19, 26, 27]. However, the performance achieved is still lower than expectation. The key reason is that the photocurrent is usually degraded in the short-wavelength region, leading to non-optimal overall performance [9, 19, 26, 27]. This phenomenon is believed to be from the Fano effect [28, 29], according to [10]. In fact, the incident solar energy experiences metallic absorption in almost the whole spectrum before reaching the active layers, further limiting the device performance. It has been shown that NPs in various metals exhibit quite different behaviors in solar light scattering and detrimental metallic absorption [30]. Indeed, we have recently carried out a comparative study in which we designed and fabricated gallium arsenide (GaAs) photodiodes in plasmonic designs with various metals, i.e., gold (Au), silver (Ag), and aluminum (Al), by metal-organic vapor phase epitaxy (MOVPE) and electron beam lithography (EBL) [31]. Optoelectronic characterization demonstrated that Al NPs outperform the other two metals in broadband photocurrent enhancement. We showed that the metallic loss can be mitigated depending upon the metal type and explained this in terms of the plasma frequency of the scattering media and the resulting field penetration into the particles [31].

In order to gain further insight into this experimental observation and seek a universal design guideline for (broadband and low loss) nanoscale scattering enhanced solar cell systems, we take a general approach in this paper to understand the balance between scattering and absorption of NPs using Mie and quasi-static calculations [4, 32]. Applying the design guideline has enabled a detailed understanding of our previous experimental result [31] and the development of an optimal design for GaAs SCs with front NPs of typical metals (Ag, Au, and Al). This optimization was carried out using our three-dimensional (3D) solar cell model, which treats both electromagnetic and carrier transport/recombination processes in detail [26, 27]. Our results show, in agreement with experiments, that SCs with front Al NPs exhibit a much higher performance compared to Au and Ag systems and show comparable performance to those configured with NPs of an 'optimal' refractive index determined from Mie calculations. This is because the refractive index of Al satisfies the criteria deduced from Mie calculation so that parasitic absorption can be dramatically suppressed while sustaining sufficiently high scattering enhancement. The design guideline opens a way to achieve high-performance nanostructured SCs with scattering NPs.

## 2. Control of scattering and absorption of NPs

The optical scattering and parasitic absorption of front NPs atop SCs are closely related to the response of each particle. Our study thus starts with a stand-alone NP in air in order to understand the underlying physics and find the way of controlling the scattering and absorption properties through adjusting the refractive index ($N = n + ik$) and physical dimensions of the particle. Mie results [32] are shown in Fig. 1 for typical radii, where (a) & (c) are for $R = 10$ nm, (b) & (d) are for $R = 80$ nm, and $<Q_{sca}>$ and $<Q_{abs}>$ are respectively the spectrally averaged scattering and absorption efficiencies [4]. The wavelength range considered is 300 nm $\leq \lambda \leq$ 1200 nm, typical for single-junction photovoltaic devices [33]. We observe that the strongest scattering and absorption appear at a small $n$ and $k \sim 1.4$. The absorption efficiency decreases rapidly with a slight deviation of $k$, while strong scattering efficiency can be sustained even under a relatively large $k$, as can be seen in Fig. 1(b) for $R = 80$ nm. A higher value for $n$ also reduces the parasitic absorption, but at a much lower rate, especially for large NPs [see Fig. 1(d) where a strong absorption region exists at $n > 3$ and $k < 4$]. Moreover, a high $n$ leads to greatly degraded NP scattering [Fig. 1(a) and Fig. 1(b)] due to the deviation from the resonant condition and therefore is not a good choice for light-trapping applications.

Accordingly, an initial conclusion can be made: the strongest scattering for light-trapping is normally accompanied by the strongest parasitic absorption due to the excitation of plasmonic resonances, which counteract the positive effect of strong scattering and limit the cell performance. Considering that small NPs contribute extremely low scattering and high absorption, a suitably large NP should be used. For large NPs, however, the balance between scattering and parasitic absorption seems to be especially crucial since the strong absorption is found in a large range of $N$. In order to obtain a high $Q_{sca}$ and an extremely low $Q_{abs}$, NP materials with a low $n$ and a high $k$ are desired (i.e., top-left corners of $n$-$k$ plane, see Fig. 1).

The dependences of $Q_{sca}$ and $Q_{abs}$ spectra on $k$ are plotted in Fig. 2 for $R = 10$ nm, where $n = 0.01$ is used in order to clearly view the dramatic changes of absorption property under a small variation of $k$ as shown in Fig. 1. Depending upon the value of $k$ we observe a number of scattering/absorption peaks in the spectra, which originate from resonant nanoparticle plasmon modes of different orders. In the plot these modes are marked as dipolar (D), quadrupolar (Q) and hexapolar (H) and we note that as the wavelength is increased high order modes vanish, leaving the dipolar mode dominant. High-order modes also become important in the region with small *k* values but within much narrower bands. Keeping photovoltaics in mind however, the dipole mode determines the scattering and absorption of such NPs, since the solar spectrum drops off rapidly at short wavelengths where the high-order resonances are prevalent. Therefore the maxima in Fig. 1(a) and Fig. 1(c) correspond to the dipolar plasmon resonance only. This also explains the distinct off-resonance behaviors shown in Fig. 1(a) and Fig. 1(c). The cross-sectional patterns of radial electric component $|E_r|^2$ of typical modes are inserted into Fig. 2(b). It is shown that for small NPs the field is strongly confined inside the particle (i.e., a strong light-NP coupling), especially for low-order plasmonic resonances, leading to a strong parasitic absorption.

Corresponding $Q_{sca}$ and $Q_{abs}$ spectra, plotted as a function of *k*, are shown in Fig. 3 for *R* = 80 nm, which takes into account a wider spectrum range to cover the red-shifted and broadened resonances. More plasmonic resonant bands with significantly enhanced $Q_{sca}$ and $Q_{abs}$ are seen. Besides the dipole resonance, quadrapole and hexapole modes have to be considered in the case of large NPs for the design of SCs since their red-shifted resonances fall into the solar spectral band ($\lambda > 300$ nm) [34]. This reveals that the unique observations from Fig. 1(b) and Fig. 1(d) are the results of hybrid modes composed by several plasmonic resonances.

Fig. 2 and Fig. 3 have also shown us that $Q_{abs}$ can be substantially suppressed by using a large *k* and a low *n*, while sustaining a relatively large $Q_{sca}$. This corresponds to the top-left corner in each subfigure of Fig. 1 and offers a promising design space for photovoltaics. The spectra of $Q_{sca}$ and $Q_{abs}$ are inserted in Fig. 3(a) by setting *N* = 0.01 + 4i and *R* = 80 nm as an example. It is observed that the broadband enhancement of scattering can be obtained and most importantly $Q_{abs} \sim 0$. Fig. 2 has shown that the strong NP absorption arises from a strong light-NP coupling, which is seen again for the large NP (see case A in Fig. 3). However, slightly increasing *k* to 2 (case B), the light-NP coupling is dramatically suppressed, showing a greatly reduced $Q_{abs}$. Further increasing *k* (case C), NP becomes "inaccessible" and contributes to scattering of the incident light without parasitic loss. It should be emphasized that *n* does not need to be zero (i.e., a lossless material since *n* = 0 gives the imaginary part of permittivity Im($\varepsilon$) = 0) for a zero absorption. As shown in the figures, a high *k* and a relatively small *n* (with noticeable material absorption) can easily contribute an extremely low metallic loss. Therefore, the resulting low absorption originates from shifting the plasmonic resonance instead of eliminating the intrinsic absorbing capability of the material.

## 3. Quasi-static analysis

The plasmonic resonating characteristics can be reflected by the NP polarizability ($\alpha_0$) obtained analytically. For discussion convenience, we define the polarizability per unit volume as $\alpha = \alpha_0/V$ ($V = 4\pi R^3/3$). In quasi-static treatment, $\alpha$ of a small NP with dimension much shorter than the wavelength (i.e., $\alpha_s$) can be written as [32]

$$\alpha_S = \frac{3(\varepsilon - \varepsilon_m)}{\varepsilon + 2\varepsilon_m} \quad (1)$$

where $\varepsilon = \varepsilon_1 + i\varepsilon_2 = N^2$ and $\varepsilon_m = 1$. While for a large NP, $\alpha_L$ is approximated by [35, 36]

$$\alpha_L = \frac{1 - (1/10)(\varepsilon + \varepsilon_m)x^2 + O(x^4)}{\left(\frac{1}{3} + \frac{\varepsilon_m}{\varepsilon - \varepsilon_m}\right) - \frac{1}{30}(\varepsilon + 10\varepsilon_m)x^2 - i\frac{4\pi \varepsilon_m^{3/2}}{3}\frac{V}{\lambda_0^3} + O(x^4)} \quad (2)$$

As $Q_{sca} \propto |\alpha|^2$ and $Q_{abs} \propto \text{Im}(\alpha)$ [4], we simply discuss $|\alpha|^2$ and Im($\alpha$) instead of $Q_{sca}$ and $Q_{abs}$.

Plots of $|\alpha|^2$, Im($\alpha$) and Re($\alpha$) versus *n* and *k* are displayed in Fig. 4 for NPs with *R* = 10 nm and *R* = 80 nm calculated from Eq. (1) and Eq. (2), respectively. Furthermore, $\lambda = 500$ nm is used as it does not modify the results qualitatively for such a non-dispersive material. The good agreement between Fig. 4(a) & Fig. 4(c) with Fig. 1(a) & Fig. 1(c) allows us to use Eq. (1) to explain the Mie-findings for $Q_{sca}$ and $Q_{abs}$ for small NPs. According to Eq. (1), the dipolar plasmonic resonance occurs at $\varepsilon_1 = -2$, i.e., $k \sim 2^{1/2}$ and $n \sim 0$, yielding $|\alpha|^2$ and Im($\alpha$) ($Q_{sca}$ and $Q_{abs}$) peaks [see (a) and (c) in Fig. 1 and Fig. 4; also noted that Im($\alpha$) = 0 at *n* = 0]. Fixing *n* and slightly deviating *k* from the resonance, Im($\alpha$) is dramatically decreased, corresponding to the suppressed $Q_{abs}$. However, a relatively high $|\alpha|^2$ can be sustained due to the presence of a large Re($\alpha$) [Fig. 4(e)], corresponding to a high scattering. This verifies that the scattering and absorption properties of small NPs shown in Fig. 1(a) and 1(c) are indeed from the dipolar resonance. However, discrepancies between quasi-static and Mie calculations emerge when considering a large particle, that is, the strong scattering region at $n > 2$ and $k \sim 0$ (dielectric NPs) and strong absorption region at $n > 3$ (high-order plasmonic modes) observed in Fig. 1(b) and Fig. 1(d)] are invisible in Fig. 4(b) and Fig. 4(d). This is due to the fact that Eq. (2) does not include the necessary high-order resonant modes [35, 36].

The parasitic absorption can also be examined in terms of the light-particle coupling. Here, the electric field inside the NP, $E_{in}$, can be obtained analytically. For example, for small particles, $E_{in}$ can be written quasi-statically as [4]

$$E_{in} = \left(1 - \frac{\varepsilon_1 - 1 + i\varepsilon_2}{\varepsilon_1 + 2 + i\varepsilon_2}\right) E_0 \quad (3)$$

where $E_0$ is the incident electric field and the 2$^{nd}$ term in the bracket denotes the depolarizing effect of electric dipoles. When $\varepsilon_2 \sim 0$ and decreasing $\varepsilon_1$ from $-2$ (corresponds to the case of $n \sim 0$ and increasing $k$ from $2^{1/2}$), $E_{in}$ goes down rapidly and approaches 0 for large $|\varepsilon_1|$. In this case, the internal built-in electric field due to the generated electric dipoles completely compensates $E_0$ [see also case C of Fig. 3(b)] and therefore the NPs show zero absorption. However, as indicated previously, $\varepsilon_2$ of zero is not a requirement for low absorption, since according to Eq. (3) $E_{in} \sim 0$ as long as $\varepsilon_1 \gg \varepsilon_2$ (as well as $\gg 2$ which is always true for metals), i.e., the left-top corners shown in Fig. 1.

## 4. Implementation in plasmonic solar cells

Next we turn our attention to the application of such scattering nanoparticles to the front surface of heterojunction GaAs SCs. The detailed device configurations, including material type, device dimension, and doping type/concentration, have been given in Fig. 5 [37]. The remaining optical and electronic parameters, which are necessary for a complete device simulation, can be found from [26, 27] and references therein.

For our first calculations we assume that the value of $N$ for the front NPs can be adjusted arbitrarily for comparison with Mie calculation. A complete evaluation of the designed SCs should be based on both optical and electrical properties. We take the short-circuit current density $J_{sc}$, which is calculated through a device-oriented 3D simulation [26, 27], as the evaluation criterion for the front nanopatterned GaAs SCs. The calculated $J_{sc}$ is plotted in Fig. 6(a), where the NPs are configured as $R = 80$ nm and $\Lambda = 400$ nm. Correspondingly, Fig. 6(b) displays the map of NP absorption percentage $P_{abs}$ (spectrally averaged), which is obtained by spatially integrating the Poynting vector over the whole particle. It is apparent that the actual $J_{sc}$ ($P_{abs}$) in a solar cell structure configured with coupled-NPs exhibits similar behavior to $Q_{sca}$ ($Q_{abs}$) of a stand-alone NP, showing the importance of the resonant characteristics of each NP and allowing us to optimize the performance of these type of SCs through controlling the scattering and absorption properties of the NP unit. In correspondence to the detrimental absorption loss, $J_{sc}$ shows the reverse response, i.e., the lowest $J_{sc}$ is resulted from the strongest plasmonic absorption (dipole resonance at $n \sim 0$ and $k \sim 2^{1/2}$). Moving away from the resonance by using a small $n$ and a large $k$ as introduced in previous sections, a large $Q_{sca}$ is sustained with negligible NP absorption; therefore, a high $J_{sc}$ is obtained. According to Fig. 6, $J_{sc}$ can be $> 21$ mA/cm$^2$ (an increase $> 4$ mA/cm$^2$ compared to a cell without NPs).

Building on these observations enables detailed examination and understanding of the performance of GaAs SCs with NPs of different metals on the front surface, in this case Au, Ag, and Al. Recent reports described optical simulations for Si-based SCs [30] and experimental measurements on GaAs devices comparing the performance of these three metals [31]. The primary conclusions of these reports indicate that Al performs better than the other metals since the blue-shifted resonance reduces NP absorption in the relevant spectral region [30, 31]. Here we focus on comparing the electrical response of GaAs-based SCs and elucidate the underlying physics.

Fig. 7 shows the $J_{sc}$ maps of the front-nanopatterned GaAs SCs under different configurations of $R$ and $\Lambda$. It should be stressed that the $J_{sc}$ value for each $R$-$\Lambda$ pair is obtained from one 3D optical and electrical calculation to acquire accurate results. Fixing $\Lambda$ and increasing $R$ from 0 to $\Lambda$, the cell is actually changing from one extreme condition to the opposite, i.e., from system 1 without NP ($R = 0$ nm) to system 2 with very high surface NP coverage ($R = \Lambda$). System 1 gives the reference $J_{sc} \sim 17$ mA/cm$^2$ and system 2 is with a strong surface reflection leading to $J_{sc} \sim 0$ mA/cm$^2$ as shown in

Fig. 7. When varying Λ under a given *R*, similar behaviors can also be found. Therefore, optimization can be realized by properly designing *R* and Λ in accordance with Fig. 7. We note that compared to Au or Ag, the Al NP system provides a much higher $J_{sc}$, agreeing well with optical predictions [38, 39] and our experiment [31].

Plotted in Fig. 8 are the corresponding EQE curves for the flat (reference) and nano-patterned cells. When using Au or Ag NPs we observe photocurrent enhancement (degradation) in long-λ (short-λ) region in accordance with previous reports [9, 19]. In fact, optical calculations of the NP absorption indicate a very high loss of the incident light when using Ag or Au due to the excitation of plasmonic resonances (not shown here, see ref [31] for example). However, Al NPs greatly improve the cell performance across almost the whole spectrum, showing a significantly enhanced light-conversion capability. We also plot the EQE response of a GaAs SC with NPs of *N* = 0.1 + 4i which falls into the optimal region in *n-k* plane. Compared to Al system, the device shows a modest improvement in performance when λ > 400 nm, contributing a similar $J_{sc}$ after spectral integration.

## 5. Conclusion

We systematically investigated the scattering and parasitic absorption of metallic nanoparticles applied to the front surface of SCs. Despite the fact that scattering can be enhanced by exciting plasmonic resonances, they are always accompanied with a strong parasitic absorption loss, which becomes a key detrimental factor limiting the cell performance. To minimize this parasitic loss, while maintaining high scattering we proposed optimized designs for the refractive index and dimensions of the NPs so that the device electrical output can be maximized. Our results showed that the desired particle refractive index should have a small real part and a large imaginary part. Among the available metals, Al has proven to be very effective in efficient light trapping since its optical constants closely satisfy the requirement. Finally, based on these observations we used our device-oriented 3D model to design a GaAs SC with Al NPs that exhibits full-band photocurrent enhancement over the reference.

**Acknowledgement**

This work is supported by National Natural Science Foundation of China (No. 61204066, No. 91233119), EU FP7 project "PRIMA" – 248154, "Thousand Young Talents Program" of China, and Priority Academic Program Development (PAPD) of Jiangsu Higher Education Institutions.

**Figures**

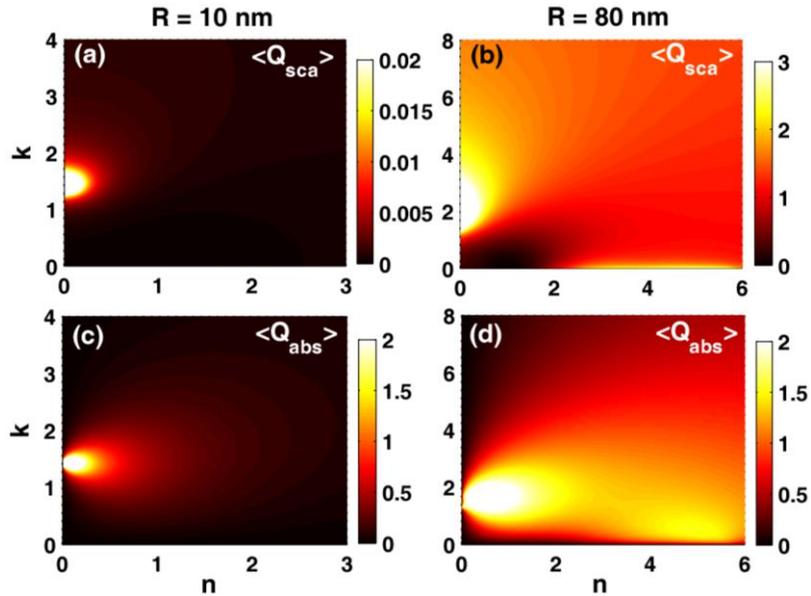

**Figure 1**: Mie-calculated <$Q_{sca}$> and <$Q_{abs}$> (spectrally averaged) versus $n$ and $k$ of a stand-alone NP in air. Small NP with $R$ = 10 nm [(a) and (c)] and large NP with $R$ = 80 nm [(b) and (d)] are considered.

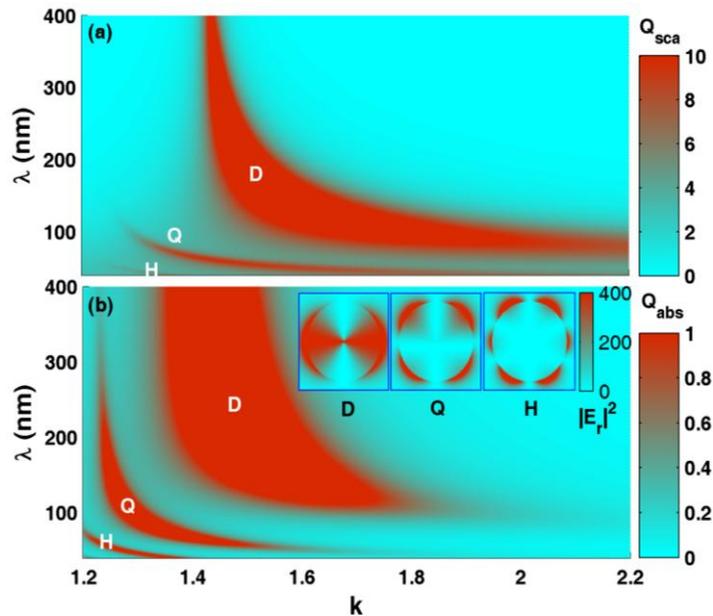

**Figure 2**: Mie-calculated $Q_{sca}$ and $Q_{abs}$ versus $\lambda$ and $k$, where $R$ = 10 nm. The insets of (b) plot the cross-sectional patterns of $r$-component electric field $|E_r|^2$ for different resonant modes. D: dipole with $N$ = 0.01 + 1.4i at $\lambda$ = 300 nm; Q: quadrapole with $N$ = 0.01 + 1.24i at $\lambda$ = 200 nm; H: hexapole with $N$ = 0.01 + 1.23i at $\lambda$ = 60 nm.

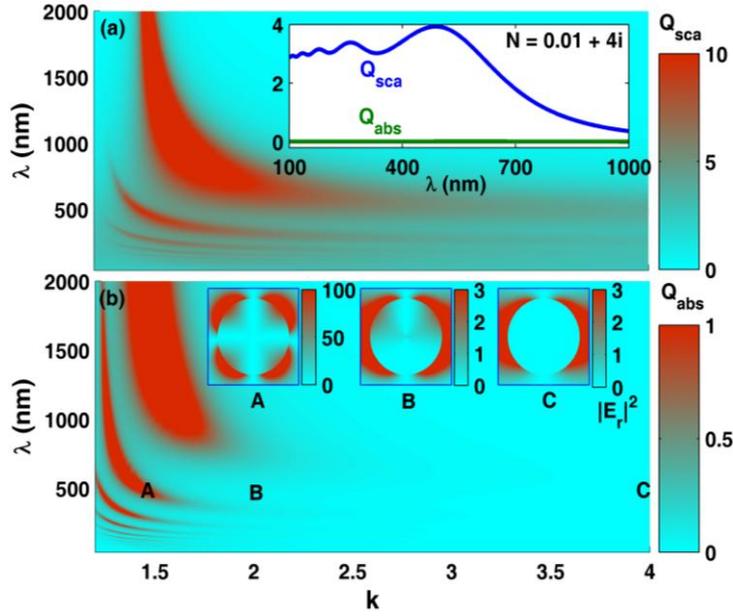

**Figure 3**: Mie-calculated $Q_{sca}$ and $Q_{abs}$ versus $\lambda$ and $k$, where $R = 80$ nm. The insets of (b) plot the cross-sectional patterns of $r$-component electric field $|E_r|^2$ for different conditions, i.e., A ($N = 0.01 + 1.4i$ at $\lambda = 500$ nm) for quadrapole resonant case with strong parasitic absorption of NP, B ($N = 0.01 + 2i$ at $\lambda = 500$ nm) for strong scattering with very low NP absorption, and C ($N = 0.01 + 4i$ at $\lambda = 500$ nm) for off-resonant case with extremely low absorption and sufficiently strong scattering. The spectra of $Q_{sca}$ and $Q_{abs}$ for the case of $N = 0.01 + 4i$ are inserted into (a).

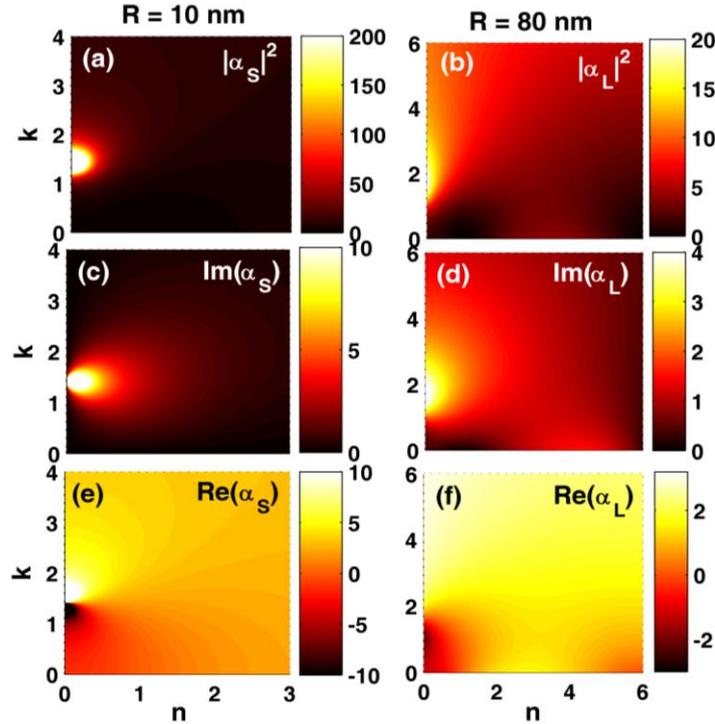

**Figure 4**: (a), (c) and (e): quasi-static calculation of $\alpha_S$ with $R = 10$ nm; (b), (d) and (f): large-particle approximation calculation of $\alpha_L$ with $R = 80$ nm. Light wavelength $\lambda$ is taken to be 500 nm.

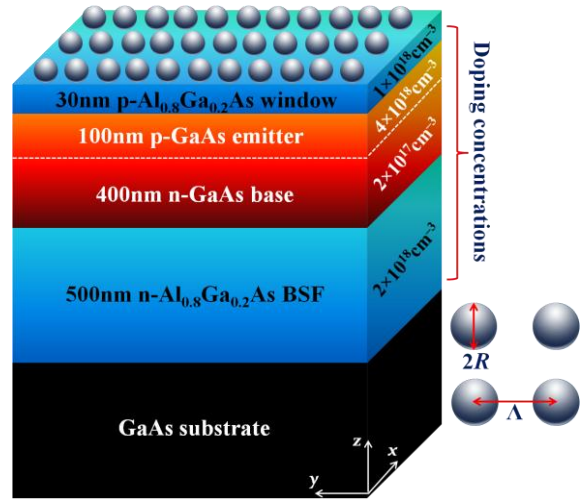

**Figure 5**: A heterojunction GaAs solar cell with front NPs of radius *R* and array period Λ.

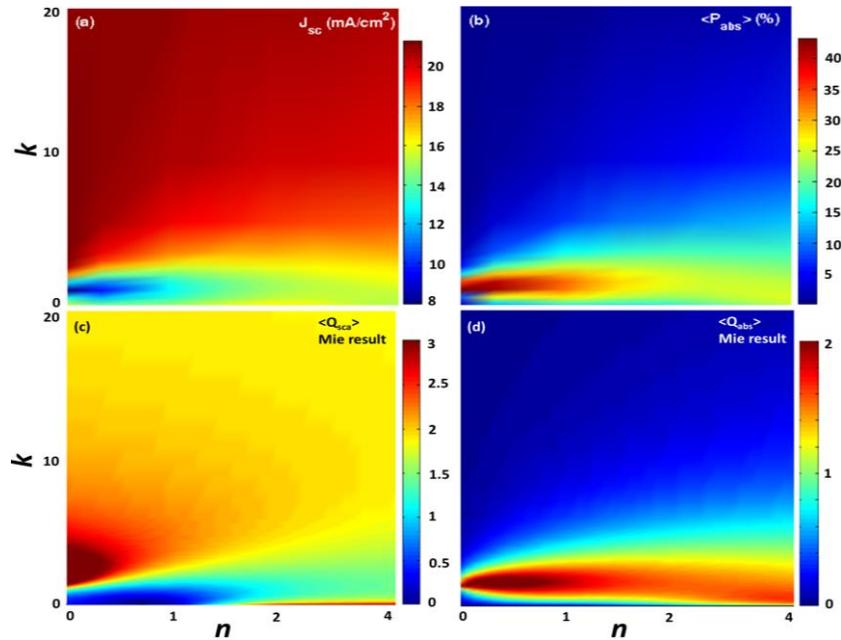

**Figure 6**: Maps of photocurrent $J_{sc}$ (a) and averaged metallic absorption $P_{abs}$ (b) with respect to *n* and *k*, showing good agreement with Mie prediction (the insets). System configuration shown in Fig. 5 with $R = 80$ nm & $\Lambda = 400$ nm is used.

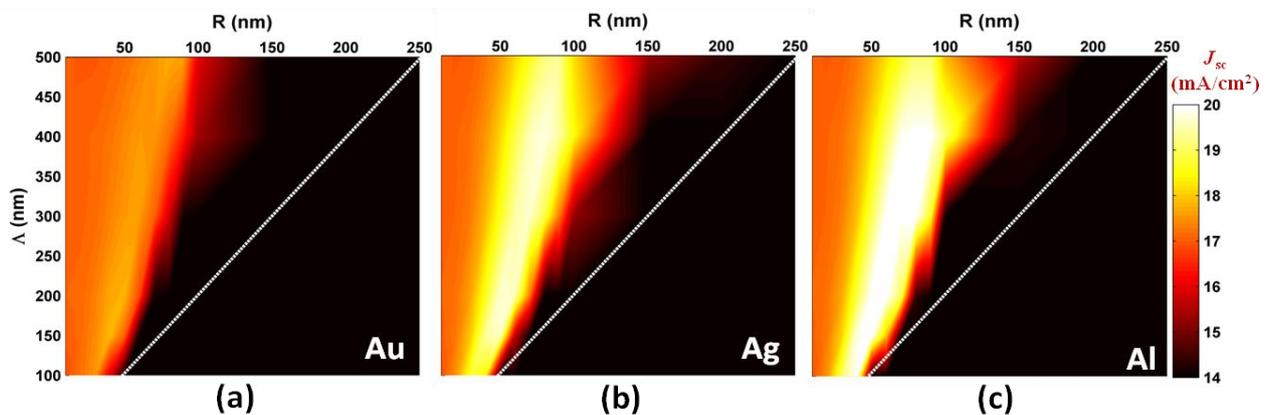

**Figure 7**: Photocurrent ($J_{sc}$) maps with respect to *R* and Λ, where Au (a), Ag (b), and Al (c) NPs are considered. Images are obtained through polynomial fitting the datum from a number of 3D calculations.

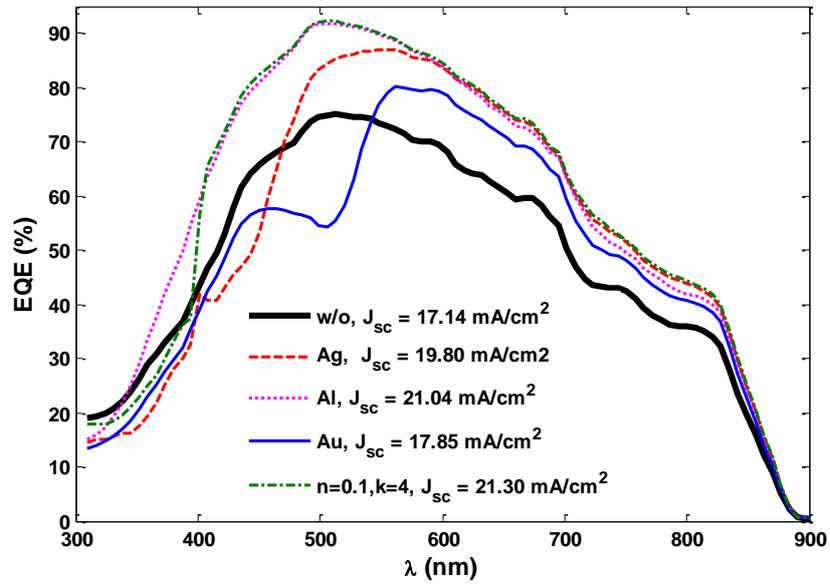

**Figure 8**: EQE responses of the optimal GaAs SCs under various NPs. The optimal configurations for GaAs SCs with Au, Ag, and Al NPs are $R = 40$ nm & $\Lambda = 150$ nm, $R = 80$ nm & $\Lambda = 400$ nm, and $R = 50$ nm & $\Lambda = 150$ nm, respectively. The original flat case (i.e., w/o) and the design with $N = 0.1 + 4i$ ($R = 80$ nm and $\Lambda = 400$ nm) are also given for comparison. The corresponding $J_{sc}$ values are also inserted.